\documentclass[sigconf]{acmart}  

\copyrightyear{2024}
\acmYear{2024}
\setcopyright{acmlicensed}

\acmConference[LLM for Individuals, Groups, and Society \@ SIGIR 2024]{Proceedings of the 47th ACM SIGIR Conference on Research and Development in Information Retrieval}{July 14--18, 2024}{Washington D.C., USA}
\acmBooktitle{Proceedings of the 47th ACM SIGIR Conference on Research and Development in Information Retrieval (SIGIR '24), July 14--18, 2024, Washington D.C., USA}

\AtBeginDocument{%
  \providecommand\BibTeX{{%
    \normalfont B\kern-0.5em{\scshape i\kern-0.25em b}\kern-0.8em\TeX}}}

\usepackage{hyperref}       
\usepackage{url}            

\usepackage{nicefrac}       
\usepackage{microtype}      
\usepackage{algorithm}  

\usepackage{algpseudocode}
\usepackage{graphicx}
\usepackage{enumitem}

\usepackage{multirow}
\usepackage{float}

\begin{document}

\title{Session Context Embedding for Intent Understanding in Product Search}

\author{Navid Mehrdad}
\affiliation{%
  \institution{Walmart Global Technology}
  \city{Sunnyvale}
  \country{USA}
}
\email{navid.mehrdad@walmart.com}

\author{Vishal Rathi}
\affiliation{%
  \institution{Walmart Global Technology}
  \city{Sunnyvale}
  \country{USA}
}
\email{vishal.rathi@walmart.com}

\author{Sravanthi Rajanala}
\affiliation{
  \institution{Walmart Global Technology}
  \city{Sunnyvale}
  \country{USA}
}
\email{sravanthi.rajanala@walmart.com}

\renewcommand{\shortauthors}{Mehrdad et al.}

\begin{abstract}
It is often noted that single query-item pair relevance training in search does not capture the customer intent. User intent can be better deduced from a series of engagements (Clicks, ATCs, Orders) in a given search session. We propose a novel method for vectorizing session context for capturing and utilizing context in retrieval and rerank. In the runtime, session embedding is an alternative to query embedding, saved and updated after each request in the session, it can be used for retrieval and ranking. We outline session embedding's solution to session-based intent understanding and its architecture, the background to this line of thought in search and recommendation, detail the methodologies implemented, and finally present the results of an implementation of session embedding for query product type classification. We demonstrate improvements over strategies ignoring session context in the runtime for user intent understanding.
\end{abstract}



\maketitle

\section{Introduction}
\label{sec:intro}

How can the current query-item pair be extended to capture the search context? And what are relative gains from incorporating search history in a fleeting search session? In this draft we answer this question by adding session context containing previous queries and engaged items into the input of embedding models, and use such augmented session state vectors for tasks such as query's associated ``product type" classification. We employ large language models (LLMs) for vectorizing the session data. In contrast to the limiting dichotomy of query item pair, we show that extending concept of query to include previous query search items can enhance user intent understanding via an accurate follow up of customer journey in a given session. We show considerable improvement in performance measures such as f1 scores of query's product type intent classification, and demonstrate that the size of gain over current query classifications for user intent is contingent on the nature of prior queries: when previous queries are broad and current query is narrow, the inclusion of the previous query, alongside with the current query, results in higher gains in user intent understanding. In contrast, training with previous--current query pairs that are of narrow--broad transitions do not yield such gains in performance. 

\section{Related Work}

\label{sec:related}

User session data has been extensively used to infer user intent, the sequence of engaged items in a session, in particular, has been used to contextualize user trajectory in a given session and to surface the next most relevant item for the user. As such, session information in recommender systems is primarily used for incorporating prior engaged {\it items} into the current recommendation \cite{core22}\cite{hidasi16}\cite{hidasi18}\cite{Seol22} \cite{sun19}\cite{Wu17}\cite{Kang18}\cite{Cho20}\cite{Tang18}. 

Instead of focusing on merely predicting next item and its attributes using session data on sequence of engaged products \cite{m223}, we examine the items engaged in a user session alongside with the {\it queries} used by the customer throughout the session. This focus on the query sequence in a given session distinguishes this study from the prior art on recommendation based on previous item engagement. 

The embeddings of items engaged in sessions can be combined to contextualize the search and recommendation experience \cite{core22}. In parallel, we introduce session embedding as an augmentation of the query embedding with previous queries and item engagements, and provide it in the runtime as a contextualized means of query understanding. Our work is distinct from query rewrite solutions \cite{Zuo23} in that we use previous queries as an addition to the current query for generating input to the transformer-based language models \cite{transformer} trained on session data.

\section{Session Embedding}
\label{sec:embed}
A typical session flow is the aggregation of query entries and item engagements, that is linked to a specific user, and is limited in time. Our design is informed by the fact that the sequence of queries and corresponding Clicks, Add To Carts (ATCs), and Orders are reliable indicators of user intent which are lost in training rerank and retrieval on {\it aggregate} query-item pair relations. Query understanding, in particular, due to its runtime implementation in the e-commerce platform in question, is expected to significantly benefit from session-based data.

Session embedding is a method for vectorization of the prior query issuance and item engagement. A variety of Large language Models (LLM)s can be used to produce such vecotrizations for inferring query attributes, reranking and retrieval. As such, session embeddings are vehicles for translating textual input from past queries, and their engaged item attributes such as item title, gender, size, brand and description into an embedding vector, which in turn can be used for user intent understanding. This vector can be used for {\it classification} as well as other tasks, this is the primary usage we have implemented the session embedding for here. Similarly, session embedding methods can be used for generative purposes alike. The session embedding vector is saved as a part of runtime infrastructure and can be accessed efficiently during the query/intent understanding process. 

Session embedding is an upgrade of query embeddings in the runtime, it can be combined with query embedding in the runtime or used independently. Session embedding is inferred using a subset of 1) previous queries, 2) previous engaged items' attributes. Engagement items include: items ordered, items clicked, items ATCed. Other customer context parameters such as facets applies, geographical and device-dependent usage information can also be added to the vectorization context. In our design, for runtime considerations, session embedding training and inference is done with a light weight language model, such as DeBERTa \cite{he2023debertav}.

During session embedding model training, the query data augmented with previous queries and previously engaged items' attributes. It is necessary to note that training labels are session based. For example in the case of query product type classification, for each successful (converted) session, labels are the converted items' product types. Training labels are based on session outcomes such as conversion, non-abandonment, Add to Cart, instead of query-item aggregate engagements.

\begin{figure*}
    \centering
    \includegraphics[width=1\linewidth]{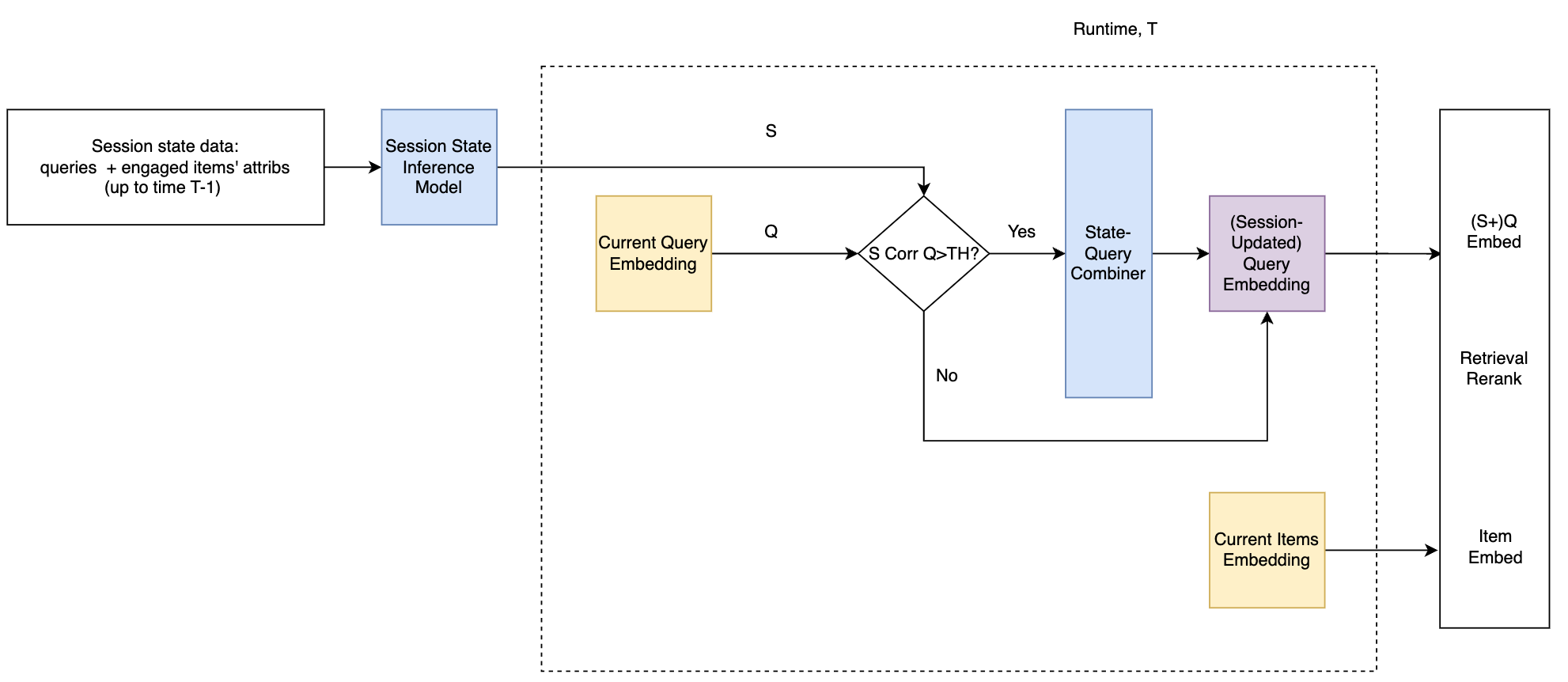}
    \caption{Session embedding}
    \label{fig:session}
\end{figure*}

A schematic of session state inference and incorporation in the runtime is included in figure (\ref{fig:session}). The previous query is linked to the current one and used for session embedding inference if there is a token match between the previous and current queries. The users can change focus during sessions, and the token match criterion acts as a guardrail against linking consequent queries that are irrelevant to each other. 

\section{Session Data} 
\label{sec:data}

During the runtime inference, session embedding is combined with query embedding in the runtime, or can completely replace it. As mentioned in the previous section, session embedding is conducted using training features included in Table (\ref{tab:embed_features}). The event-based nature of the labels is in contrast to that of a host of prior vectorization methods for search in production that are produced on the aggregate query-item level. The aggregate labels are averages in time of the levels of engagement between the query and item. Instead we have opted for labels that are session-based.

\begin{table}
    \centering
    \begin{tabular}{|l|}
    \hline
     previous quer(ies) \\ \hline 
    previous engaged items' attributes \\ 
    \ \ \ \ \ -- items ordered    \\ 
      \ \ \ \ \ --  items ATCed  \\ 
     \ \ \ \ \  -- items clicked  \\ \hline
     facets applied   \\ \hline
     geo, device\\
     \hline
    \end{tabular}
    \caption{Data used in session embedding}
    \label{tab:embed_features}
\end{table}

The state of the session ($S_T$) is a function of past queries and past engaged items' attributes up to time T-1. The state is inferred prior to runtime inferences at time T. At runtime, the current query embedding Q, is combined with the session vector S so that an augmented query embedding (Q+S) can be used for session-aware retrieval and rerank.

Session up to time T-1 is combined with current query only if a minimum correlation between $S_{T-1}$ and $Q_T$ exists. While the generalized flow of session embedding in the following can use a session state vector and current query vector for correlation measurement, we have used simpler correlation mechanisms such as pre-vectorization {\it token match} between previous query $Q_{T-1} and Q_T$ current query. If the state is taken as simply previous query and the state-query combination is a simple concatenation, then session embedding boils down to an LLM inference on $(Q_{T-1},Q_T)$. In the evaluation section, we show that such a simple scheme is capable of major improvement over the existing query product type classifier in production. Most importantly, the state (e.g. previous query) should be used in training and inference, only if the correlation between session state and the current query is above a certain threshold. This ensures cases of multiple intent in a given session are not counted as an indicator of continuous intent. 

\subsection{Training data}
Session data from three months of September, October and November in 2023, previous and current queries $(q_{i-1}, q_i)$ pairs with $q_i$ having at least one order, while $q_{i-1}$ lacking orders. These query pairs are filtered to produce pairs with token matches. Labels are the product type of ordered items for $q_i$ . 

Training dataset contains 44.7M data points. In addition to previous queries, we have built datasets in which session state includes previous query's ATCed and clicked item attributes. 


\subsection{Queries' inferred product types}
The primary use case for the session embeddings we have trained is to infer the product types of current queries. Examples of $(q_{i-1}, q_i)$ and their corresponding product types are included in table (\ref{tab:pts}). 

\begin{table*}
    \centering
    \begin{tabular}{|l|l|}
    \hline
 Previous Query, Current Query  & Previous Query PT, Current Query PT\\ \hline\hline
        ['thermal lunch box school', 'thermal lunch box'] & ['Reusable Lunch Bags \& Boxes', 'Reusable Lunch Bags \& Boxes'] \\ \hline
        ['bottled water', 'gallon water'] & ['Bottled Drinking Waters', 'Bottled Drinking Waters']\\ \hline
        ['spaghetti sauce', 'spaghetti noodles'] & ['Pasta Sauces', 'Pasta']\\ \hline
         ['guinea pig cages', 'guinea playpen']&['Small Animal Habitats \& Cages', 'Small Animal Habitats \& Cages'] \\ \hline
         ['contact lens solution', 'saline solution'] &['Contact Lens Cleaners', 'Contact Lens Cleaners'] \\ \hline
         ['pool shock', 'algaecide for pools'] & ['Pool Chemicals', 'Pool Chemicals']\\ \hline
         ['celsius', 'celsius mix in'] & ['Energy Drinks', 'Drink Mixes']\\ \hline
    \end{tabular}
    \caption{Current and previous queries with product types}
    \label{tab:pts}
\end{table*}

\section{Query-Item Relevance in Sessions}
\label{sec:methods}

Augmenting the query and item to the session level implies queries can be considered as a combination of search terms, taken into account together, not separately. The same applies to the items in the session. 

To elucidate the point about (item1,item2) being surfaced for (query1,query2) pair, on the session level, and based on engagement label training, one can imagine (item1,item2) as an augmentation of individual items, and (query1,query2)  as an augmentation of query. If we knew that (query1,query2) occurs, we are well-positioned to surface (item1,item2). $item_{i-1},item_i$ for $query_{i-1},query_i$, the same way that surface $item_i$ is surfaced for $query_i$. 

The occurrence of $(,,. q_{i-k}, ..., q_{i-2}, q_{i-1}, q_i, .., )$, and their engaged/converted items, is sequential, so we are building a state vector $S_i$ that can represent  $(,,. q_{i-k}, ..., q_{i-2}, q_{i-1})$ up to time i, well beforehand and combine that with the vectorized version of the present query for a seamless state integration.

 \begin{table*}
    \centering
    \begin{tabular}{l|c|c}
     & {\bf f1 on test (weighted, 6k+ class)} &{\bf \# datapoints training}   \\ \hline \hline
     current query, previous query, previous ATCed item attribs   & 85.14\%	 & 6,346,447  \\ \hline
     current query, previous query, previous Clicked item attribs   & 83.62\%	& 4,811,928   \\ \hline
     current query, previous query: {\bf only broad to narrow transitions} &  {\bf 85.42}\%	& 12,605,589  \\ \hline
     current query, previous query: only narrow to broad transitions    &  80.38\%	& 5,648,265   \\ \hline
    current query, previous query  &  83.72\% &	44,671,909   \\ \hline
    current query  & 82.92\%	& 44,671,909    \\ \hline
    \end{tabular}
    \caption{Training Results}
    \label{tab:session_results}
\end{table*}

\section{Results \& Discussion}
\label{sec:models_train}

We examined including varieties of prior session activity in addition to the current query for the goal of user intent understanding. Focusing on query's implied product type, to be chosen from a set of multiple thousands, we trained light weight LLMs for the classification task. Our goal was to re-enact the user experience as closely as possible in order to yield a better performance. For that purpose, we experimented with including 1) previous query with token match 2) previous narrow queries transitioning to broad queries, and 3) previous broad queries transitioning to narrow queries. In other words, queries in 1) were filtered for either 2) narrow to broad or 3) broad to narrow transitions only. 

Including the previous query (with token match) improved the weighted f1 score for the 6k+ classes of query product types, see Table (\ref{tab:session_results}). 

To probe further, we filtered the training data based on the nature of query transitions from broad to narrow or vice versa. The number of broad to narrow transitions in the dataset was about 12.6M, narrow to broad 5.6M, from a total of 44.6M query transitions between queries with token match. 

Interestingly enough, the performance, measured by weighted f1 score, deteriorated when we tried training on narrow to broad query transitions. This is surprising, as narrow queries were expected to include attributes that are key to retrieving relevant items for the next broad queries in the session. Nevertheless, we see that while there may be cases with that pattern in our dataset, it is more likely that the customers already achieved the results intended by the attributes in previous queries and their use of a current broad query, and associated order, is a sign of failed focused search, and hence may not be closely related to the current query's broad intent. 

However, when including only broad to narrow transitions, we find major improvement of about 2\% in weighted f1 score over indiscriminate usage of query transitions and about 2.5\% over usage of current query only. We ascribe this improvement in performance to the match between the user journey and the nature of broad-narrow transitions. It is likely that users' search funnel for successful sessions starts with broad queries and narrows down as users zoom into the items they intend to choose. This outcome is against the common , and immediate, perception that previous queries' main use is through their narrow attributes pertaining to the current query. We find the opposite to be true. 

Finally, we also included attributes from items 1) clicked and 2) ATCed (but not ordered) from the previous query in the training dataset. While the addition of the previously clicked items' attributes did not help the performance, having previous ATCed items' attributes considerably improved the outcome of the training. The numbers are included in Table (\ref{tab:session_results}).

Results in table (\ref{tab:session_results}) show considerable improvement over training query's product type (PT) intent classifier only on the current query. Most important observation is that while broad--narrow query transition pairs are highly beneficial to session-based training, the narrow-broad ones are not. 

We have used a light weight LLM (DeBERTa V3 small) to train the query product type classifier with previous query as session state. Larger models can also be employed, but runtime consideration needed to be taken into account. 

For calculating the f1 scores we used the maximum odds produced over the 6k+ classes of product types for each query inferred. We examined using a odds threshold for inferring a class of product types instead. The number of inferred product types per query, for our test set of 50k queries, are included in (\ref{tab:odds}). Among the 6k+ class odds predicted on the test set data points, majority of them are concentrated in one or two product types and the maximum spread of product types inferred is 6.

\begin{table}[!htbp]
    \centering
    \begin{tabular}{c|l}
        \# of predicted types & frequency \\ \hline \hline
        0 & 148\\ \hline
        1 & 35873\\ \hline
        2 & 11101\\ \hline
        3 & 2480\\ \hline
        4 & 370\\ \hline
        5 & 27\\ \hline
        6 & 1\\ \hline \hline
    
        Total & N=50k \\ 
    \end{tabular}
    \caption{Distribution of the number of predicted product types with odds ratio > 0.1 in the t test set, total $N=50k$ }
    \label{tab:odds}
\end{table}

\section{Conclusion and Future Work}
The results demonstrate the utility of including session history in user intent understanding. We showed that such utility is contingent on the careful usage of query evolution trajectory for training. Broad to narrow query transitions, mapping to the user search funnel, effectively improve the performance of query product type classification. The findings have implications for both classificatory and generative tasks of intent understanding in the search setup. Leveraging efficient large language models, for vectorizing the session space, are key to productionalizing such session-based user intent understanding designs.

\small 
\bibliographystyle{ACM-Reference-Format}
\bibliography{references}

\end{document}